\documentclass[preprint,12pt]{elsarticle}

\usepackage{amssymb}

\usepackage{epsfig}

\usepackage{appendix}

\journal{Nuclear Physics A}

\newcommand{\beq}{\begin{eqnarray}}
\newcommand{\eeq}{\end{eqnarray}}

\begin{document}

\begin{frontmatter}

\title {Sensitivity of  $\Lambda$ single-particle energies to
the $\Lambda N$ spin-orbit coupling and to nuclear core 
structure in p-shell and sd-shell hypernuclei}

\author{P. \textsc{Vesel\'y}$^{1}$, E. \textsc{Hiyama}$^{2}$, J. \textsc{Hrt\'{a}nkov\'{a}}$^{1}$ and J. \textsc{Mare\v{s}}$^{1}$}

\address{$^{1}$Nuclear Physics Institute, Czech Academy of Sciences,
250 68 \v{R}e\v{z}, Czech Republic \\
$^{2}$RIKEN Nishina Center, RIKEN, Wako, Saitama 351-0198, Japan}

\begin{abstract}
We introduce a mean field model based on realistic 2-body baryon interactions 
and calculate spectra of a set of $p$-shell and $sd$-shell $\Lambda$ 
hypernuclei - $^{13}_{\Lambda}$C, $^{17}_{\Lambda}$O, $^{21}_{\Lambda}$Ne, 
$^{29}_{\Lambda}$Si and $^{41}_{\Lambda}$Ca. The hypernuclear spectra are 
compared with the results of a relativistic mean field (RMF) model and 
available experimental data. The sensitivity of $\Lambda$ single-particle 
energies to the nuclear core structure is explored. Special attention is 
paid to the effect of spin-orbit $\Lambda N$ interaction on the energy 
splitting of the $\Lambda$ single particle levels $0p_{3/2}$ and $0p_{1/2}$. 
In particular, we analyze the contribution of the symmetric (SLS) and the 
anti-symmetric (ALS) spin-orbit terms to the energy splitting. We give 
qualitative predictions for the calculated hypernuclei. 
\end{abstract}

\begin{keyword}
$\Lambda$ hypernuclei \sep  spin orbit splitting \sep $\Lambda N$ interacion \sep 
mean field model


\end{keyword}

\end{frontmatter}

\section{Introduction}

One of the major goals in hypernuclear physics is to obtain information on 
baryon-baryon interactions in a unified way. However, due to the limitation 
of hyperon ($Y$) nucleon ($N$) scattering data, the $YN$ potential models 
proposed so far, such as the Nijmegen models, have a large degree of 
ambiguity. Therefore, quantitative analyses of light $\Lambda$ hypernuclei, 
where the features of $\Lambda N$ interactions appear rather clearly in 
observed level structures are indispensable. For this purpose, 
accurate measurements of 
$\gamma$-ray spectra \cite{Tamura00,Ajimura01,Akikawa02,Ukai04,Ukai06} 
and high resolution $(\pi^+,K^+)$ reaction experiment \cite{Hotchi01} 
have been performed systematically. These experiments are source of invaluable 
information about spin-dependent components of $\Lambda N$ interaction. 
Useful constraints on the $YN$ interaction components have been provided 
by shell model \cite{Millener85} and few-body cluster \cite{Hiyama2009,Hiyama2012} calculations. 
 Among the $YN$ spin-dependent components, spin-orbit terms are important, since they are 
intimately related to modeling the short-range part of the interaction. For example, it is well 
known that the antisymmetric spin-orbit (ALS) forces come out qualitatively 
different in one-boson-exchange (OBE) models 
\cite{model-D,model-F,soft-core,Rijken99} 
and in quark models \cite{Morimatsu84,QCM1}. 
It was pointed out that the ALS force based on quark model \cite{QCM1} 
is as strong as to cancel the symmetric spin-orbit (SLS) force, while the 
ALS force of Nijmegen potentials \cite{model-D,model-F,soft-core,Rijken99} 
is of a smaller strength. To extract information on these spin-orbit forces, 
specific $\gamma$-ray experiments \cite{Ajimura01,Akikawa02} and 
$(\pi^+, K^+)$ reaction experiments \cite{Hotchi01} have been performed.
In the $\gamma$-ray experiments, spin-orbit dominated energy splittings 
of $43 \pm 5$ keV for the $3/2^+_1$ and $5/2^+_1$ doublet of levels in 
$^9_{\Lambda}$Be \cite{Akikawa02} and $152 \pm 54 \pm 36$ keV for the 
$1/2^-_1 - 3/2^-_1$ doublet in $^{13}_{\Lambda}$C \cite{Ajimura01} were 
measured. Considerably larger energy splittings of $1.37 \pm 0.20$, 
$1.63 \pm 0.14$ and $1.70 \pm 0.10 \pm 0.10$ MeV were reported from the 
analysis of the $p$-orbit, $d$-orbit and $f$-orbit peaks, respectively, 
observed in the $^{89}_{\Lambda}$Y spectrum of the $(\pi^+, K^+)$ reaction 
\cite{Hotchi01}. However, these splittings are most likely caused by 
core-excited configurations and have little to do with the spin-orbit 
interaction~\cite{Motoba2008}. 

The strength of the spin-orbit forces was studied within 
few-body \cite{Hiyama2000} and shell model calculations \cite{Millener2012} 
of $^9_{\Lambda}$Be and $^{13}_{\Lambda}$C. Reasonable strengths were 
introduced to reproduce the experimental data. It is worth mentioning that 
small energy splittings for $^9_{\Lambda}$Be and $^{13}_{\Lambda}$C were 
predicted by Hiyama et al. \cite{Hiyama2000} using ALS forces based on a quark 
model. On the other hand, it was pointed out in the shell-model calculation 
by Millener \cite{Millener2001} that a $\Lambda N$ tensor contribution was 
as important as spin-orbit for these energy splittings. 

More information about spin-orbit interaction in other $\Lambda$ hypernuclei 
is certainly needed. For this purpose, it is planned to explore the level structure 
of some medium heavy $\Lambda$ hypernuclei at JLab. Our aim in the present paper 
is to study the spin-orbit doublet states $p_{1/2}$ and $p_{3/2}$ in $\Lambda$ 
hypernuclei built on $n \alpha$ nuclear cores. It is expected that the 
spin-spin and tensor terms of the $\Lambda N$ interaction are not effective 
as a consequence of the $\alpha$-cluster dominating structure, and we can thus safely 
assess $YN$ spin-orbit forces. However, if $\alpha$ 
clusters in these hypernuclei are broken, the spin-spin and the 
tensor interactions might contribute to the $p_{1/2}$ and $p_{3/2}$ 
energy splitting as well. In this paper, we investigate spin-spin and spin-orbit 
contributions to this energy splitting and give qualitative predictions of the splittings 
in $^{13}_{\Lambda}$C, $^{17}_{\Lambda}$O, $^{21}_{\Lambda}$Ne, $^{29}_{\Lambda}$Si and 
$^{41}_{\Lambda}$Ca. For this aim we apply the ESC08c potential, which has 
been recently proposed by Nijmegen group, using a mean field approach based 
on realistic baryon interactions \cite{HYPproc}. In the next section, we 
briefly describe our method. Selected results are presented and discussed 
in the third section, and a summary is given in the final section. 

\section{Methodology}

We calculate spectra of single $\Lambda$ hypernuclei using a Hamiltonian of the form
\begin{equation}
H = T + V^{NN} + V^{\Lambda N} - T_{\rm{CM}} , 
\label{Ham1}
\end{equation}
where $T=\sum^A_{a=1} p_a^2/2m$ is the kinetic term, $V^{NN}$ ($V^{\Lambda N}$) denotes interaction among nucleons ($\Lambda$ and nucleons) and $T_{\rm{CM}} = (\sum^A_{a=1} p_a^2 + \sum^A_{a\ne b} \vec{p}_a.\vec{p}_b)/2mA$ is a center of mass term. In our calculations we use a realistic $\Lambda N$ interaction and a realistic $NN$ interaction corrected by the density dependent term which simulates $3N$ force. If we introduce the creation (annihilation) operators $a^{\dagger}_i$ ($a_i$) for nucleons and $c^{\dagger}_i$ ($c_i$) for $\Lambda$, we can express the Hamiltonian (\ref{Ham1}) in the formalism of second quantization
\begin{equation}
H = \sum_{ij} t^{N}_{ij} a^{\dagger}_i a_j + \sum_{ij} t^{\Lambda}_{ij} c^{\dagger}_i c_j + \frac{1}{4} \sum_{ijkl} V^{NN}_{ijkl} a^{\dagger}_i a^{\dagger}_j a_l a_k + \sum_{ijkl} V^{\Lambda N}_{ijkl} a^{\dagger}_i c^{\dagger}_j c_l a_k ,
\label{Ham2}
\end{equation}
with the kinetic matrix elements 
\begin{equation}
t_{ij}= \left( 1 - \frac{1}{A} \right) \langle i|\frac{p^2}{2m}|j\rangle ,
\label{kin1}
\end{equation}
antisymmetrized $NN$ interaction matrix elements 
\begin{equation}
V^{NN}_{ijkl}=\langle ij|\left( V^{NN} - \frac{\vec{p}_1.\vec{p}_2}{2mA} \right)|kl-lk\rangle ,
\label{int1}
\end{equation}
and $\Lambda N$ interaction matrix elements 
\begin{equation}
V^{\Lambda N}_{ijkl}=\langle ij|\left( V^{\Lambda N} - \frac{\vec{p}_1.\vec{p}_2}{2mA} \right) |kl \rangle ,
\label{int2}
\end{equation}
all expressed in the harmonic oscillator basis. The harmonic oscillator basis depends on one parameter $\hbar \omega_{\rm{HO}}$ which defines the oscillator lengths $b_{N}$ and $b_{\Lambda}$ for the wave functions of nucleons and $\Lambda$, respectively, due to the relations
\begin{equation}
b_{N} = \sqrt{\frac{\hbar^2c^2}{m_{N}c^2 \hbar\omega_{\rm{HO}}}} , \hspace{2cm} b_{\Lambda} = \sqrt{\frac{\hbar^2c^2}{m_{\Lambda}c^2 \hbar\omega_{\rm{HO}}}} .
\label{blength}
\end{equation}
In basis which is large enough the physical results do not depend on $\hbar \omega_{\rm{HO}}$. In our calculations we use $N_{max} = 10$ and $\hbar \omega_{\rm{HO}} = 16$ MeV.

The mean field is constructed in our model as follows. First we solve the nuclear part of Hamiltonian (\ref{Ham2}) $(\sum_i t^{N}_{ij} a^{\dagger}_i a_i + \frac{1}{4} \sum_{ijkl} V^{NN}_{ijkl} a^{\dagger}_i a^{\dagger}_j a_l a_k)$ within the Hartree-Fock approximation. As a result we obtain the wave function $|{\rm HF}\rangle$ as well as the nuclear density 
$\rho^{N}_{lk}= \langle {\rm HF} |a^{\dagger}_k a_l| {\rm HF} \rangle$ of the nuclear core of a hypernucleus. Then we calculate the single-particle energies $e^{\Lambda}_i$ by diagonalizing the matrix $(t^{\Lambda}_{ij} + U^{\Lambda N}_{ij})$ where
\begin{equation}
U^{\Lambda N}_{ij} = \sum_{\tau=p,n} \sum_{kl} V^{\Lambda N}_{kilj} \rho^{N}_{lk} ,
\label{U_field}
\end{equation}
assuming that the $\Lambda$ hyperon interacts with the mean field of the core nucleus. The hypernuclear wave function at the level of mean-field approximation is defined as $|i\rangle = c^{\dagger}_i 
|{\rm HF}\rangle$.

In our model, we used the realistic $NN$ interaction NNLO$_{\rm{opt}}$~\cite{n2lo}. It is a chiral next-to-next-to leading order potential with parameters optimized to minimize the effect of three-body interactions (their effect, however, still remains relatively important), which makes this force useful for many-body calculations (for more details about the optimization procedure see \cite{n2lo}). Matrix elements of this interaction were generated by the CENS code \cite{Hjort}. However, the effect of three-body forces is still not negligible. If we perform the calculations purely with the two-body $NN$ interactions we do not obtain correct distribution of the nuclear density. The nuclear density distribution is more compressed which leads to much smaller nuclear rms radii than are the experimental values. 
In general, this has influence on the single particle energies of $\Lambda$, particularly on the splitting between 
the $0s$ and $0p$ states as can be deduced from the Bertlmann-Martin inequalities \cite{Bertl}. For this reason we add a corrective density dependent (DD) term of the form
\begin{equation}
V^{NN,DD} = \frac{C_{\rho}}{6} (1 + P_{\sigma}) \rho\left( \frac{\vec{r}_1+\vec{r}_2}{2} \right) \delta(\vec{r}_1-\vec{r}_2) ,
\label{DDterm}
\end{equation}
where $C_{\rho}$ is the coupling constant and $P_{\sigma}$ is the spin exchange operator. This density dependent interaction term was first introduced in \cite{Hergert}. It was shown \cite{Waroq} that it gives the same contribution to the Hartree-Fock energy as the contact three-body interaction
\begin{equation}
V^{NNN} = C_{\rho} \delta(\vec{r}_1-\vec{r}_2) \delta(\vec{r}_2-\vec{r}_3).
\label{contact3b}
\end{equation}
The term (\ref{DDterm}) is necessary for reasonable description of correct nuclear single particle spectra within 
the mean field calculations with realistic $NN$ interactions \cite{Ves} and improves significantly the description 
of the nuclear density distributions and radii \cite{Ves2}.
In this paper we fix the values of $C_{\rho}$ for each hypernucleus independently to get the realistic values of nuclear radii as well as nuclear densities with respect to the available experimental data \cite{Radii} and the calculations within the RMF model \cite{Jarka}.   
Our future goal is to implement directly the chiral $NNN$ interaction instead of the density dependent term (\ref{DDterm}). In this case we would not need to fit any independent parameter $C_{\rho}$. 

The $\Lambda N$ interaction implemented in our model is the YNG force derived from the Nijmegen model ESC08 \cite{esc}, namely its version ESC08c \cite{Isaka}. It is represented in a three-range Gaussian form:
\begin{equation}
G(r;k_{\rm{F}}) = \sum_{i=1}^{3} (a_i + b_i k_F + c_i k_F^2) \mbox{exp}(-r^2/\beta_i^2).
\label{Gaussian}
\end{equation}
For more details including the values of the parameters $a_i$, $b_i$, $c_i$, $\beta_i$ see~\cite{Isaka}. We represent the 
$\Lambda N$ interaction (\ref{Gaussian}) in the form of the interaction elements of Eq.~(\ref{int2}).
It should be noted that there are no tensor terms in the ESC08c version \cite{Isaka} used in this work.

The $\Lambda N$ interaction depends explicitly on the Fermi momentum $k_{\rm{F}}$. We can either consider 
$k_{\rm{F}}$ as a free parameter of our model and fit its value to the observed hypernuclear spectra or we can fix 
the value of $k_{\rm{F}}$ within the Thomas-Fermi approximation through the relation
\begin{equation}
k_{\rm{F}} = \left( \frac{3 \pi^2}{2} \langle \rho\rangle \right)^{1/3} .
\label{ThomasFermi}
\end{equation}
The average density $\langle \rho \rangle$ in Eq.~(\ref{ThomasFermi}) can be expressed within the Average Density Approximation (ADA) by the following prescription \cite{ADA} 
\begin{equation}
<\rho> = \int \mbox{d}^3r  \rho_N(\vec{r}) \rho_{\Lambda}(\vec{r}) ,
\label{avrho}
\end{equation}
where $\rho_N(\vec{r})$ is the density of the nuclear core and $\rho_{\Lambda}(\vec{r})$ is the density of $\Lambda$ in the hypernucleus.  Note that we have to perform the hypernuclear calculation to obtain $\rho_{\Lambda}(\vec{r})$ and determine the value of Fermi momentum $k_{\rm{F}}$. In other words the value of $k_{\rm{F}}$ has to be evaluated self-consistently because the equation (\ref{avrho}) depends on the result of a calculation which itself depends on $k_{\rm{F}}$. 

The symmetric (SLS) and antisymmetric (ALS) spin-orbit terms in the $\Lambda N$ potential are included within the Scheerbaum approximation \cite{Scheer}. Due to this approximation we include the effect of the SLS and ALS terms directly at the mean field level. We add the following contribution into the matrix (\ref{U_field}): 
\begin{equation}
U^{N\Lambda,\rm{ls}}_{ij} = \langle i| K_{\Lambda} \frac{1}{r} \frac{\mbox{d}\rho}{\mbox{d}r} \vec{l}.\vec{s} |j\rangle ,
\label{Uls}
\end{equation}
where 
\begin{equation}
K_{\Lambda} = K^{\rm SLS}_{\Lambda} + K^{\rm ALS}_{\Lambda} = -\frac{\pi}{3}(S_{\rm{SLS}}+S_{\rm{ALS}}) ,
\label{KL}
\end{equation}
and
\begin{equation}
S_{\rm{SLS,ALS}} = \frac{3}{\bar{q}} \int^{\infty}_{0} r^3 j_1(\bar{q}r) G(r;k_{\rm{F}}) \mbox{d}r .
\label{SLS}
\end{equation}
The value of $\bar{q}$ in (\ref{SLS}) is set to 0.7 fm$^{-1}$ \cite{Scheer}. The form of the function $G(r;k_{\rm{F}})$ in (\ref{SLS}) is identical for the SLS and ALS terms, they only differ by the values of the input parameters.
It should be noted that the $\Lambda N$-$\Sigma N$ coupling term in ESC08c is renormalized into the $\Lambda N$-$\Lambda N$ part of G-matrix interaction, giving rise to an important part of the density dependence \cite{Isaka}. 

\bigskip

In the RMF approach, the strong interactions among point-like hadrons are mediated by \emph{effective} mesonic 
degrees of freedom. The formalism is based on the Lagrangian density of the form 
$$
{\mathcal L} = {\mathcal L}_{N} + {\mathcal L}_{\Lambda}\; ,   
$$
\begin{equation}
{\mathcal L}_{\Lambda}
=\bar{\psi}_{\Lambda}\left[{\rm i}\gamma_\mu \partial^\mu - g_{\omega \Lambda}\gamma_\mu \omega^{\mu} 
-(M_{\Lambda}+g_{\sigma \Lambda}\sigma) \right]\psi_{\Lambda} + {\mathcal L}_{\rm T} \; , 
\end{equation}
$$ 
{\mathcal L}_{\rm T} = \frac{f_{\omega \Lambda}}{2 M_{\Lambda}} 
\bar{\psi}_{\Lambda} \sigma^{\mu\nu} \partial_{\nu} \omega_{\mu} \psi_{\Lambda}\; .  
$$
Here, ${\mathcal L}_{N}$ is the standard nuclear Lagrangian \cite{Jarka} and we used the NL-SH parametrization in 
this work \cite{NLSH}. The ${\mathcal L}_{\rm T}$ is 
the $\omega \Lambda \Lambda$ anomalous (tensor) coupling term. This term is crucial in order to get negligible 
$\Lambda$ spin-orbit splitting for larger values of the $\Lambda$ couplings required by constituent quark model~\cite{jen, Mares}.

The system of coupled field equations for both baryons ($N$, $\Lambda$) and considered meson fields  
results from ${\mathcal L}$ using standard techniques and approximations \cite{Jarka,Mares}.  
 
For the coupling constants $g_{\omega \Lambda}$ and $f_{\omega\Lambda}$ we used the naive quark model 
values and $g_{\sigma \Lambda}$ was tuned so as to reproduce the binding energy of $\Lambda$ in the 
$0s$ state in $^{17}_{\Lambda}$O \cite{Mares}.

\section{Results}

In this section, we present selected results of our calculations of the hypernuclei $^{13}_{\Lambda}$C,
$^{17}_{\Lambda}$O, $^{21}_{\Lambda}$Ne, $^{29}_{\Lambda}$Si and $^{41}_{\Lambda}$Ca. In the hypernuclei with the doubly magic nuclear core - $^{17}_{\Lambda}$O and $^{41}_{\Lambda}$Ca - the calculations within Hartree-Fock method are straightforward. However, in the case of $^{13}_{\Lambda}$C, $^{21}_{\Lambda}$Ne and $^{29}_{\Lambda}$Si, the ground states of the 
corresponding core nuclei have more complex structure~\cite{Soyeur,Miller} and it is necessary to 
take into account configuration mixing and perform calculations within a deformed basis in order to describe their 
structure properly. Nevertheless, even calculations within the spherical HO basis could provide interesting information 
about these hypernuclei if we consider various configurations of the corresponding nuclear cores. 
In this work, we performed calculations for the following configurations: 
$p_{3/2}^4$ and $p_{3/2}^2 p_{1/2}^2$ in $^{12}$C;  $d_{5/2}^2$, $s_{1/2}^2$, and $d_{3/2}^2$ in  $^{20}$Ne; 
$d_{5/2}^6$, $d_{5/2}^4 s_{1/2}^2$ and $d_{5/2}^2 s_{1/2}^2 d_{3/2}^2$ in $^{28}$Si. 
We treated these configurations always symmetrically for both protons and neutrons. 
It is to be noted that more configurations could be realized in the ground states of the above nuclei. 
We selected just some of them in order to illustrate the effect of the wave function of the 
nuclear core on the energy splitting of the $\Lambda$ single particle levels $0p_{3/2}$ and $0p_{1/2}$ in the 
considered hypernuclei. 

For each particular nucleus we first fixed the parameter $C_{\rho}$ to obtain reasonable    
density distribution and rms radius of the nuclear core, comparable with the available experimental data and 
RMF calculations within the NL-SH parametrization~\cite{NLSH}. The values of the charge rms radii are 
summarized in Table~1. 

\begin{table}[b!]
\caption{The charge rms radii (in fm) of considered nuclei in selected g.s. configurations, calculated without the DD term (\ref{DDterm}), $C_{\rho}=0$~MeV$\cdot$fm$^{-6}$ (A), for the fitted values of $C_{\rho}$ (B) and within the RMF model NL-SH~\cite{NLSH} (see text for details). The experimental values (exp.) are taken from \cite{Radii}.}
\begin{center}
\label{t1}
\begin{tabular}{lllll}
\hline
  & A & B & RMF & exp. \\
\hline
$^{12}$C & 2.37 & 2.50 & 2.46 & 2.47 \\
$^{16}$O & 2.44 & 2.72 & 2.70 & 2.70 \\
$^{20}$Ne & 2.62 & 2.95 & 2.88 & 3.01 \\
$^{28}$Si & 2.73 & 3.14 & 3.04 & 3.12 \\
$^{40}$Ca & 2.99 & 3.48 & 3.45 & 3.48 \\
\hline
\end{tabular}
\end{center}
\end{table}

\begin{figure}[t!]
\includegraphics[width=30pc]{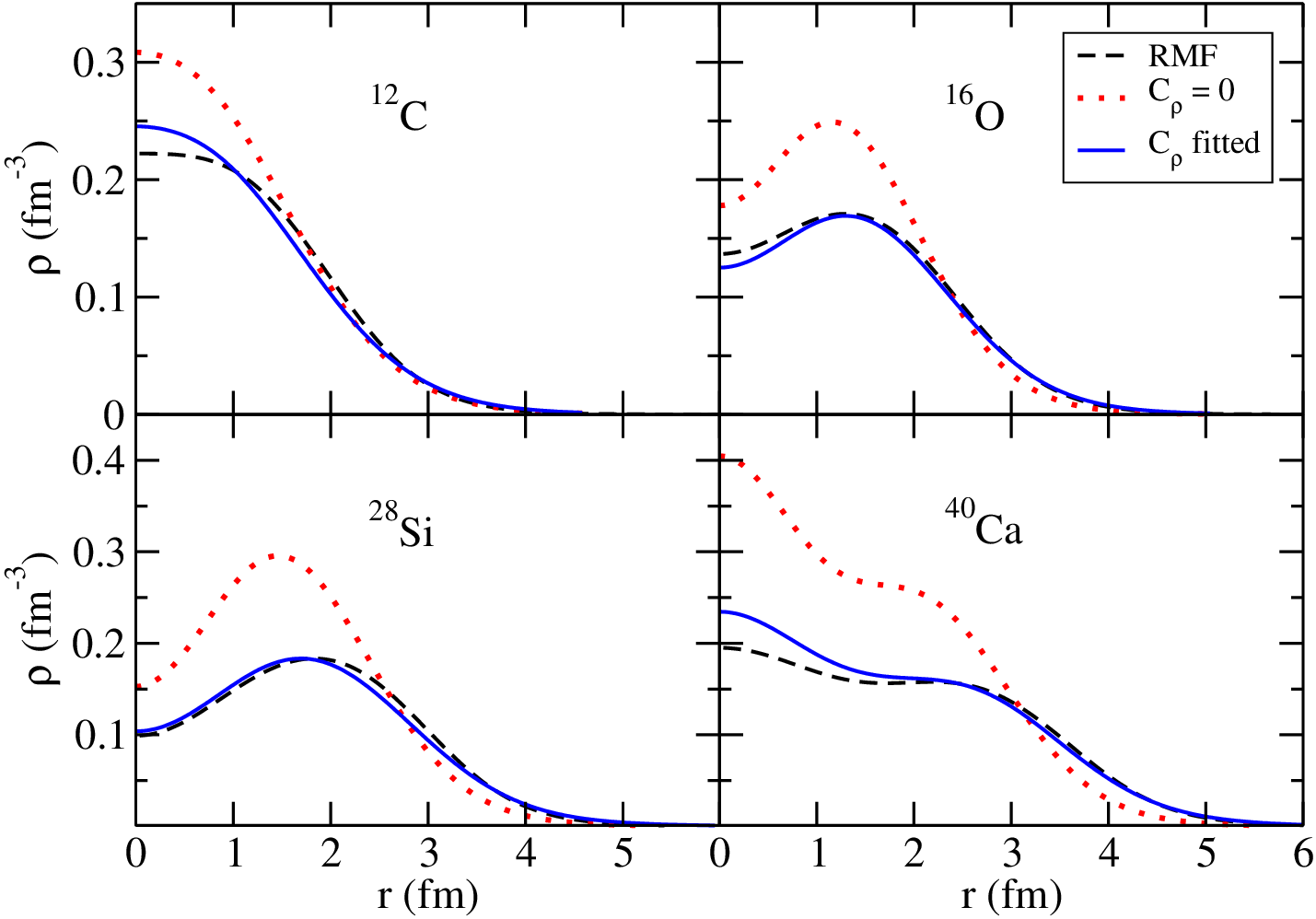}
\caption{  \label{fig1} 
The nuclear core density distributions in selected g.s. configurations of $^{12}$C, $^{16}$O, $^{28}$Si, and $^{40}$Ca, calculated without the DD term ($C_{\rho} = 0$) and with the DD term ($C_{\rho}$ fitted). 
The density distributions calculated within the RMF model NL-SH~\cite{NLSH} are shown for comparison (see text for details). 
} 
\end{figure}

We used the values $C_{\rho}=600$~MeV$\cdot$fm$^{-6}$ for $^{12}$C, $C_{\rho}=1600$~MeV$\cdot$fm$^{-6}$ for $^{16}$O, 
$C_{\rho}=1700$~MeV$\cdot$fm$^{-6}$ for $^{20}$Ne, $C_{\rho}=1600$~MeV$\cdot$fm$^{-6}$ for $^{28}$Si, and 
$C_{\rho}=2100$~MeV$\cdot$fm$^{-6}$ for $^{40}$Ca. In case of hypernuclei with the open-shell core we did not repeat  
tunning of the parameter $C_{\rho}$ for each configuration separately but we fixed it for one case (the configuration $p_{3/2}^4$ for $^{12}$C, $d_{5/2}^2$ for $^{20}$Ne and $d_{5/2}^6$ for $^{28}$Si). The corresponding radii 
for the remaining configurations differ from the values shown in Table~1 but the differences are much smaller than 
the differences between the values in the columns A and B for each given nucleus. We chose to fit the radii and radial 
distributions for the above configurations because they are the lowest configurations in energy due to the empirical 
ordering of the $s$ and $d$ levels.  

In Fig.~1,  we present the radial nuclear density distributions calculated within the mean field model 
based on realistic 2-body baryon interactions and the RMF model NL-SH~\cite{NLSH}. 
The figure illustrates the importance of the DD term in the $NN$ interaction~(\ref{DDterm}) on selected nuclei --  
$^{12}$C (in the $p_{3/2}^4$ configuration), $^{16}$O, $^{20}$Ne (in the $d_{5/2}^2$ configuration), 
$^{28}$Si ( in the $d_{5/2}^6$ configuration), and $^{40}$Ca. 
Calculations performed without the DD term 
($C_{\rho} = 0$) yield unrealistically large central densities and, as a consequence, the corresponding rms radii 
are too small. After including the DD term the density distributions become more diffused and get closer to the RMF 
predictions which are in agreement with empirical density distributions~\cite{Jarka,neon} 
(compare also charge rms radii in Table~1).


The values of the Fermi momentum $k_{\rm F}$ used in the present calculations were determined using the ADA approximation (Eqs.~(\ref{ThomasFermi}) and (\ref{avrho})). 
We obtained $k_{\rm F}=1.20$~fm$^{-1}$ for $^{13}_{\Lambda}$C (in the configuration $p_{3/2}^4$), $k_{\rm F}=1.20$~fm$^{-1}$ for $^{17}_{\Lambda}$O, $k_{\rm F}= 1.21$~fm$^{-1}$ 
for $^{21}_{\Lambda}$Ne (in the configuration $d_{5/2}^2$), $k_{\rm F}= 1.27$~fm$^{-1}$ for $^{29}_{\Lambda}$Si (in the configuration $d_{5/2}^6$), and $k_{\rm F}= 1.29$~fm$^{-1}$ for $^{41}_{\Lambda}$Ca. 

Before focusing on the main objective of the present work, the energy splittings between $p_{1/2}$ and
$p_{3/2}$, we discuss on the  $\Lambda$ single particle energies in the considered hypernuclei. 
Fig.~2 shows  our  hypernuclear spectra, calculated for the configurations for which the parameters $C_{\rho}$ and $k_F$ were tuned, and the spectra calculated within the RMF model NL-SH for the same  configurations. 
We see that our results are in good agreement with the experimental data shown for comparison. 
However, it should be pointed out that the data in the figure are for $^{16}_{\Lambda}$O, $^{28}_{\Lambda}$Si, and $^{40}_{\Lambda}$Ca since the data for hypernuclei with the same 
$Z$ and closed nuclear cores are not available at present (unlike the $^{13}_{\Lambda}$C case). 
\begin{figure}[t!]
\includegraphics[width=30pc]{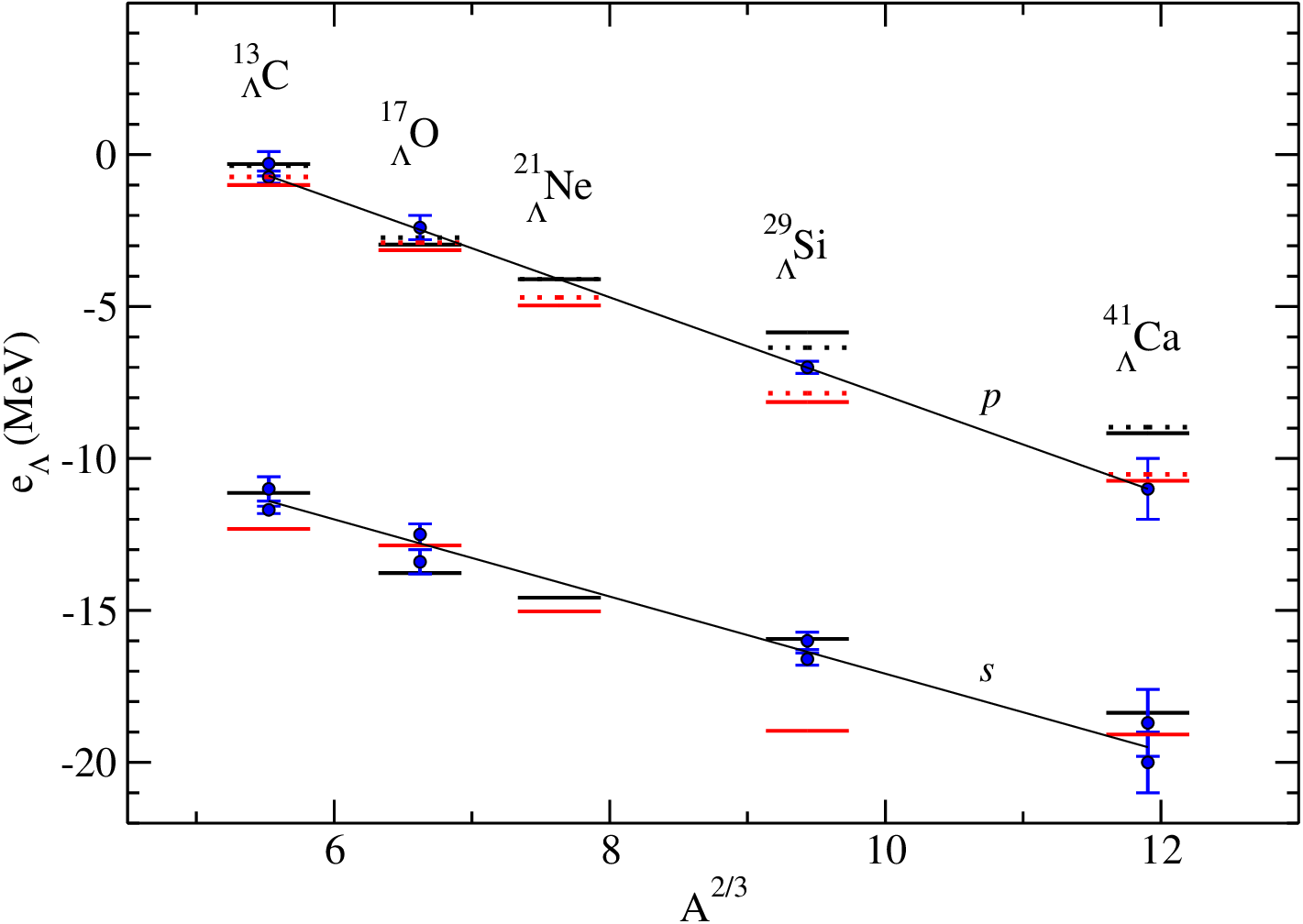}
\caption{  \label{fig2}
The $\Lambda$ single particle energies in $^{13}_{\Lambda}$C in the
nuclear core configuration $p_{3/2}^4$, $^{17}_{\Lambda}$O,
$^{21}_{\Lambda}$Ne in the configuration $d_{5/2}^2$,
 $^{29}_{\Lambda}$Si in the configuration $d_{5/2}^6$, and
$^{41}_{\Lambda}$Ca,  calculated within the mean-field model with
realistic interactions
(black lines) and the RMF model NL-SH (red lines). The $0p_{1/2}$ levels
are denoted by dotted lines.
The experimental values for $^{13}_{\Lambda}$C are taken from
\cite{expO,May}. We show also energies of the $s$ and $p$
levels measured for $^{16}_{\Lambda}$O \cite{expO,Pile}, 
$^{28}_{\Lambda}$Si \cite{Pile,Haseg} and $^{40}_{\Lambda}$Ca
\cite{Pile,Chri}.}
\end{figure}
 The most pronounced difference between the HF calculations based on realistic interactions and the RMF model NL-SH occurs in the spectrum of $^{29}_{\Lambda}$Si. Here the RMF model predicts significantly more binding for $\Lambda$ - the lowest level $0s_{1/2}$ has nearly the same energy as the $0s_{1/2}$ level in $^{41}_{\Lambda}$Ca. 
It will be demonstrated in Fig.~4 that the discrepancy between the calculated $^{29}_{\Lambda}$Si spectra is smaller for other nuclear core 
configurations and could be attributed to considerably different nuclear spin-orbit splittings in the 
two  considered models.
 We expect that proper calculations allowing for nuclear core deformation and configuration mixing of the nuclear core wave function could decrease this discrepancy in predicted 
$\Lambda$ spectra in $^{29}_{\Lambda}$Si. 
It is to be noted that the $\Lambda$ energies calculated in both considered models could get closer to each other if we fine tuned e.g. the $k_F$ parameter for each particular configuration of the nuclear core in the mean field model based on realistic interactions and/or the  RMF $g_{\sigma\Lambda}$ coupling separately for $^{29}_{\Lambda}$Si. However, being aware of the limitations of our current hypernuclear calculations we do not intend to do so in the present study. 

Since our $\Lambda$ single particle energies reproduce the data reasonably well, let us discuss on 
the main subject of our study, the energy splitting $\Delta p = [(E(0p_{1/2}) - E(0p_{3/2}))]$ between the $\Lambda$ single particle levels $0p_{3/2}$ and $0p_{1/2}$. 
In Table~2, we present the 
calculated values of $\Delta p$ for the following options: the spin-orbit (SLS+ALS) forces completely switched off (A), 
only the SLS term included~(B), both the SLS and ALS terms included (C). The results obtained within the RMF model are presented for comparison. The configurations for which the parameters $C_{\rho}$ and $k_F$ were tuned are given in bold face.
\begin{table}[t!]
\caption{The $\Lambda$ energy splitting $\Delta p = (E(0p_{1/2}) - E(0p_{3/2}))$ (in MeV) in $^{13}_{\Lambda}$C, $^{17}_{\Lambda}$O, $^{21}_{\Lambda}$Ne, $^{29}_{\Lambda}$Si, and $^{41}_{\Lambda}$Ca 
for selected nuclear core configurations, calculated without
SLS and ALS terms (A), with SLS term only (B), and with SLS + ALS terms 
(C). Negative values of $\Delta p$ indicate inverse ordering of the 
$0p_{3/2}$ and $0p_{1/2}$ levels. 
In columns (B) and (C) we present the SLS and ALS contribution to $\Delta p$, respectively. 
The values of
$\Delta p$ calculated within the RMF model are shown for comparison. Those configurations for which we tuned the parameters
$C_{\rho}$ and $k_F$ are shown in bold face. }

\begin{center}
\label{t3}
\begin{tabular}{l|l|c|cc|cc|c}
\noalign{\smallskip}\hline\noalign{\smallskip}
& core & A & \multicolumn{2}{c|}{B} & \multicolumn{2}{c|}{C}  & RMF \\ 
& configuration & $\Delta p$ & SLS &$\Delta p$ & ALS & $\Delta p$  & $\Delta p$ \\
\noalign{\smallskip}\hline\noalign{\smallskip}
$^{13}_{\Lambda}$C &  $\mathbf{p_{3/2}^4}$ & -0.58 & 1.57 & $\; 0.99$ & -1.05 & -0.06   & 0.27 \\
& $p_{3/2}^2\, p_{1/2}^2$ & $\; 0.07$ & 0.86 & $\; 0.93$  & -0.57 & $\; 0.36$ &  0.13 \\
\noalign{\smallskip}\hline\noalign{\smallskip}
$^{17}_{\Lambda}$O & $\mathbf{p_{3/2}^4\, p_{1/2}^2}$  & -0.07 & 0.89 & $\; 0.82$ & -0.59 & 
$\; 0.23$ & 0.24 \\
\noalign{\smallskip}\hline\noalign{\smallskip}
& $\mathbf{d_{5/2}^2}$ & -0.22 & 0.69 & $\; 0.47$ & -0.46 & $\; 0.01$  & 0.26 \\
$^{21}_{\Lambda}$Ne & $s_{1/2}^2$ & -0.10 & 1.28 & $\; 1.18$ & -0.85 & $\; 0.33$  & 0.25 \\
& $d_{3/2}^2$ & $\; 0.20$ & 0.56 & $\; 0.76$ & -0.37 & $\; 0.38 $ & ---  \\
\noalign{\smallskip}\hline\noalign{\smallskip}
& $\mathbf{d_{5/2}^6}$ & -0.64 & 0.39 & -0.25 & -0.25 & -0.50  & 0.29 \\
$^{29}_{\Lambda}$Si & $d_{5/2}^4 s_{1/2}^2$ & -0.51 & 1.14 & $\; 0.64$ & -0.74 & -0.10  & 0.32 \\
& $d_{5/2}^2 s_{1/2}^2 d_{3/2}^2$ & $\; 0.01$ & 1.06 & $\;1.07$ & -0.69 & $\; 0.37$ &  0.24 \\
\noalign{\smallskip}\hline\noalign{\smallskip}
$^{41}_{\Lambda}$Ca & $\mathbf{d_{5/2}^6 d_{3/2}^4 s_{1/2}^2}$ & -0.01 & 0.58 & $\; 0.57$ 
& -0.37 & $\; 0.20$  & 0.21 \\
\noalign{\smallskip}\hline\noalign{\smallskip}
\end{tabular}
\end{center}
\end{table}

It is to be stressed that the energy splittings $\Delta p$ calculated within the mean field based on realistic baryon interactions and RMF models are of different origin. In the former case, 
the splitting is caused by the 2-body $\Lambda N$ interaction with its various spin dependent terms. On the other hand, the splitting of the spin orbit partners in the RMF model results 
from a delicate balance between strong scalar and vector mean fields in the Dirac equation~\cite{jen}. This explains qualitatively 
different predictions for $\Delta p$ within the two approaches. Clearly the calculations based on 
realistic baryon interactions give considerably larger variations of the $\Delta p$ values in the studied hypernuclei than the RMF approach. 

The ESC08c $\Lambda N$ potential gives a non-zero, negative energy 
splitting of the $0p_{3/2}$ and $0p_{1/2}$ levels in several configurations in $^{13}_{\Lambda}$C, $^{21}_{\Lambda}$Ne and $^{29}_{\Lambda}$Si
 even if the SLS and ALS forces are switched off (see column A in Table~2). 
The spin-spin term in the ESC08c $\Lambda N$ interaction thus contributes significantly to the $\Delta p$ splitting 
in these configurations of hypernuclear cores.  
In the case of $^{17}_{\Lambda}$O and $^{41}_{\Lambda}$Ca (hypernuclei with the doubly magic nuclear core) and also in $^{13}_{\Lambda}$C for the configuration $p_{3/2}^2 p_{1/2}^2$,  in $^{21}_{\Lambda}$Ne for the configuration $s_{1/2}^2$ and in $^{29}_{\Lambda}$Si for the configuration $d_{5/2}^2 s_{1/2}^2 d_{3/2}^2$
the $0p_{3/2}$ and $0p_{1/2}$ levels are close to be degenerate. We checked that these states are roughly degenerate also in the other two hypernuclei with doubly magic core - $^{91}_{\Lambda}$Zr and $^{209}_{\Lambda}$Pb. 
When the SLS term is included (B), the $\Delta p$ splitting becomes positive in all considered hypernuclei except $^{29}_{\Lambda}$Si in the $d_{5/2}^6$ core configuration. 
The ALS term acts in an opposite way to the SLS term (thus weakens the effect of SLS) and its magnitude is $\approx 2/3$ of the SLS magnitude. 
The combined (SLS + ALS) term (column C) remains strong enough to cause positive $\Delta p$ splitting in most of the cases except $^{29}_{\Lambda}$Si in the core configurations $d_{5/2}^6$ and $d_{5/2}^4 s_{1/2}^2$, and $^{13}_{\Lambda}$C in the core configuration $p_{3/2}^4$. 
It is also to be pointed out that in the cases when the states $p_{3/2}$ and $p_{1/2}$ are close to be degenerate in the column A, the energy splitting $\Delta p$ comes almost entirely from the SLS and ALS spin-orbit interaction terms. 

\begin{figure}[t!]
\begin{center}
\includegraphics[width=30pc]{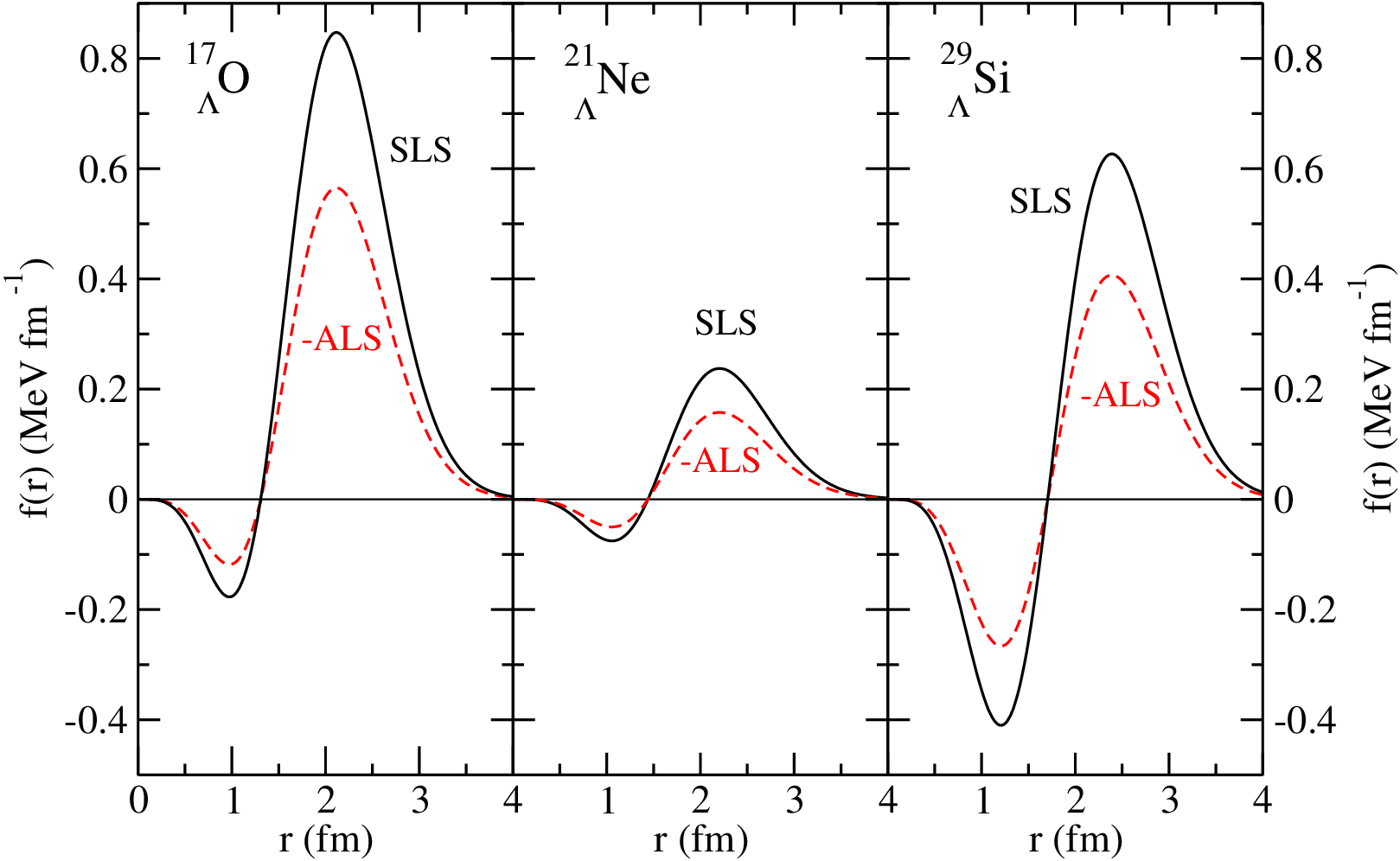}
\end{center}
\caption{ \label{fig3} 
The functions $f_{\rm SLS}(r)$ (SLS, solid line) and  $ -f_{\rm ALS}(r)$ (-ALS, dashed line) 
defined by Eq.~(\ref{func}) calculated for $^{17}_{\Lambda}$O (left), for $^{21}_{\Lambda}$Ne in the 
$d^2_{5/2}$ configuration of the nuclear core (middle), and for $^{29}_{\Lambda}$Si in the 
$d^6_{5/2}$ core configuration (right) (see text for details). 
} 
\end{figure}
Finally, there is a missing RMF value for the $^{21}_{\Lambda}$Ne configuration $d_{3/2}^2$ in Table 2 -- this configuration could not be calculated since the proton  $d_{3/2}^2$ level was found to be unbound in the applied RMF parametrization.  

The competition between the SLS and ALS forces can be illustrated with the help of the function $f(r)$ defined as follows:  
\begin{equation}
f(r) = 4\pi r^2 \phi^{*}_{0p_{1/2}}(r) K_{\Lambda} \frac{1}{r} \frac{\mbox{d}\rho}{\mbox{d}r} \vec{l}.\vec{s} \phi_{0p_{1/2}}(r) - 4\pi r^2 \phi^{*}_{0p_{3/2}}(r) K_{\Lambda} \frac{1}{r} \frac{\mbox{d}\rho}{\mbox{d}r} \vec{l}.\vec{s} \phi_{0p_{3/2}}(r), 
\label{func}
\end{equation}
evaluated for $K^{\rm SLS}_{\Lambda}$ ($K^{\rm ALS}_{\Lambda}$). The 
radial integral of $f(r)$  determines the contribution of the SLS (ALS) force to the energy splitting between the $0p_{1/2}$ and $0p_{3/2}$ levels (see Eq.~13).  

In Fig.~3, we compare $f(r)$ for $K^{\rm SLS}_{\Lambda}$ with $f(r)$ for $ - K^{\rm ALS}_{\Lambda}$  in $^{17}_{\Lambda}$O (left panel), $^{21}_{\Lambda}$Ne in the configuration $d_{5/2}^{2}$ (middle panel) and $^{29}_{\Lambda}$Si in the configuration $d_{5/2}^{6}$ (right panel).  
The difference between the areas delimited by SLS and -ALS curves above and below zero 
determines the spin-orbit (SLS + ALS) contribution to the energy splitting $\Delta p$. 
This difference is relatively large in  $^{17}_{\Lambda}$O while in the case of $^{21}_{\Lambda}$Ne, the "negative" and "positive" contributions compensate more each other. 
Even larger compensation effect is seen for the $^{29}_{\Lambda}$Si calculated within the configuration $d_{5/2}^6$. 
Consequently, the (SLS + ALS) splitting in  $^{29}_{\Lambda}$Si calculated in this configuration
is about twice smaller than in $^{17}_{\Lambda}$O (compare Table~2). 

\begin{figure}[t!]
\begin{center}
\includegraphics[width=23pc]{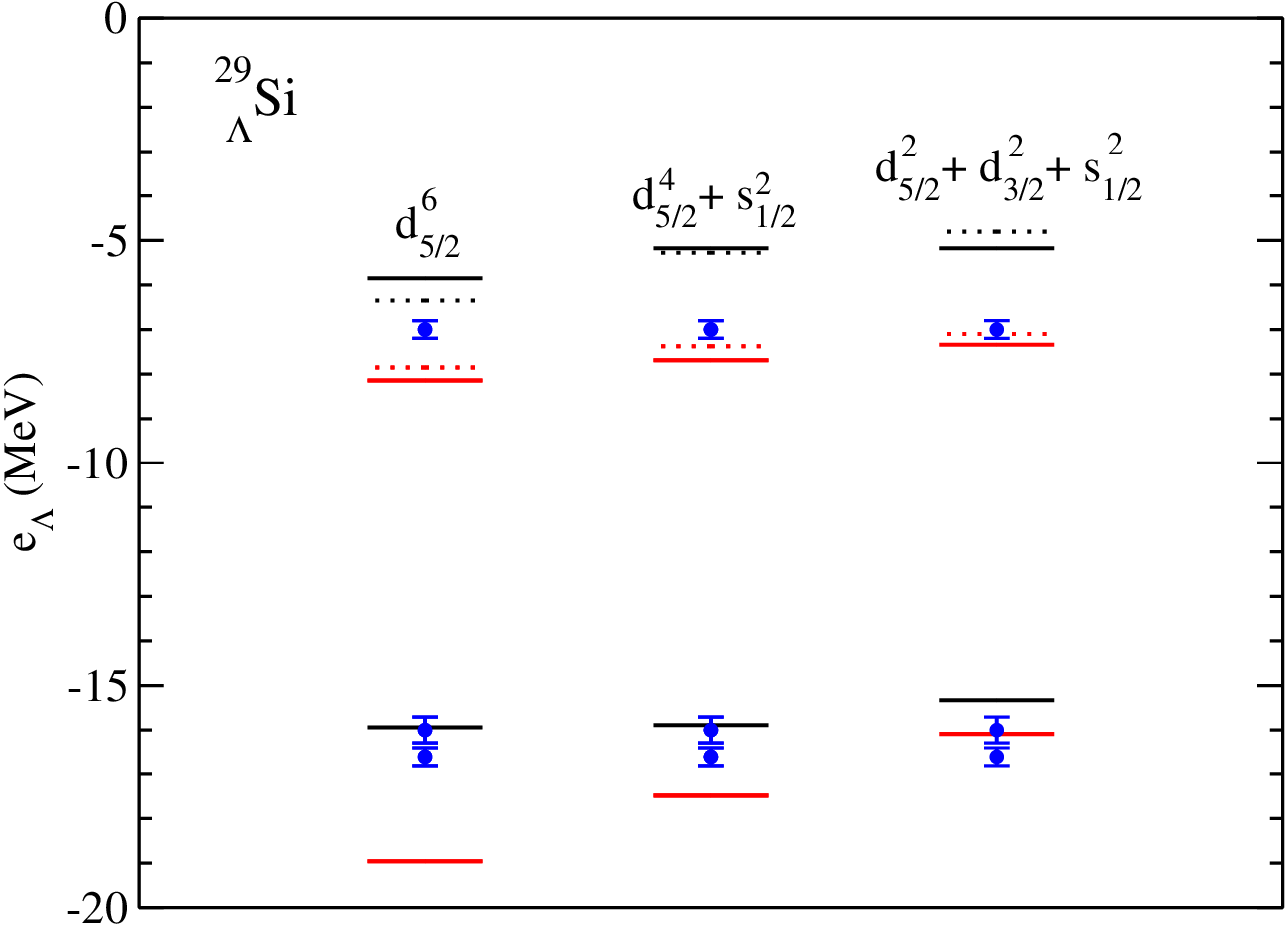}
\end{center}
\caption{  \label{fig4}
The $\Lambda$ single particle energies in
 $^{29}_{\Lambda}$Si with three different configurations of the nuclear
core, calculated within the
mean-field model with realistic interactions (black lines) and the RMF
model NL-SH (red lines). The $0p_{1/2}$ levels are denoted by dotted
lines.
The energies of the $s$ and $p$ levels measured for
$^{28}_{\Lambda}$Si are taken from \cite{Pile,Haseg}.}
\end{figure}
In Fig.~4, we show the $\Lambda$ single particle energies in
$^{29}_{\Lambda}$Si with three different configurations of the nuclear
core, calculated within the
mean-field model with realistic interactions (black lines) and the RMF
model NL-SH (red lines). The $0p_{1/2}$ levels are denoted by dotted
lines.
The experimental values for $^{28}_{\Lambda}$Si~\cite{Pile,Haseg} are shown for comparison 
(there are no data for $^{29}_{\Lambda}$Si). The figure demonstrates how the $\Lambda$ single
particle spectrum is affected by the wave function of the nuclear core. We can see that the
$\Lambda$ levels $0p_{3/2}$ and $0p_{1/2}$ switch their ordering in the 
case of mean-field based on realistic baryon forces. This is in
contrast to the RMF model for which the energy splitting of both levels
remains roughly constant. We found possible explanation for this effect by
analyzing the nucleon single particle energies. While the RMF model is able to
reproduce well the empirical spin-orbit splitting of the $0p_{3/2}$ and
$0p_{1/2}$ states for protons and neutrons, the mean field model based on
realistic baryon forces appears to be quite sensitive to different configurations considered for the ground state. We do not obtain realistic splitting of the
nucleonic $0p_{3/2}$ and $0p_{1/2}$ states in the configurations
$d_{5/2}^6$ and $d_{5/2}^4 s_{1/2}^2$ -- in the former case we even get wrong
ordering of these levels. Only for the configuration $d_{5/2}^2 s_{1/2}^2
d_{3/2}^2$ we obtain satisfactory agreement of the nucleonic spin-orbit
splitting with the empirical values (and also the RMF values). Consequently, we also get 
standard ordering of the $\Lambda$
$0p_{3/2}$ and $0p_{1/2}$ levels (see Table~2).

\section{Conclusions}

In this work, we performed calculations of selected $p$- and $sd$- shell hypernuclei, namely $^{13}_{\Lambda}$C, $^{17}_{\Lambda}$O, $^{21}_{\Lambda}$Ne, $^{29}_{\Lambda}$Si and $^{41}_{\Lambda}$Ca 
within the mean field model based on realistic 2-body baryon interactions and compared the results with 
the predictions of the RMF model NL-SH. We introduced the density dependent 2-body interaction term which mimics 
the effect of the 3-body $NNN$ force in order to get realistic charge radii and density distributions of the nuclear cores of the studied hypernuclei. 
This appeared important also in the calculations of hypernuclear spectra since the ESC08c $\Lambda N$ interaction 
depends explicitly on the Fermi momentum $k_{\rm{F}}$ which was determined using the averaged density approximation. 
Reasonable description of the density distributions in the studied (hyper)nuclear systems is thus crucial.  

The main objective of the present calculations is to study the influence
of SLS and ALS spin orbit terms on the
energy splitting $\Delta p$ of the $\Lambda$ levels $0p_{3/2}$ and
$0p_{1/2}$.
The $\Delta p$ splittings in $^{17}_{\Lambda}$O and $^{41}_{\Lambda}$Ca,
calculated within the mean field model
based on realistic baryon interactions and the RMF model NL-SH are very
close to each other.
In the case of $^{13}_{\Lambda}$C, $^{21}_{\Lambda}$Ne and
$^{29}_{\Lambda}$Si it is desirable to perform calculations within 
deformed basis and take into account configuration mixing of the nuclear core 
wave function. 
Nevertheless, our calculations in the spherical HO basis, which considered several 
configurations of the nuclear core of these hypernuclei yielded valuable insight 
into the issue of the spin dependence of the $\Lambda N $ interaction and the 
$\Lambda$ spin-orbit splitting in these open-shell hypernuclei.   
 We found that the energy splittings of the  $\Lambda$ levels
$0p_{3/2}$ and $0p_{1/2}$ calculated using realistic $NN$ and $\Lambda N$ interactions 
depend strongly on the chosen configuration of the nuclear core, unlike the RMF 
approach. For the configurations which give the energy splitting close to zero when the $\Lambda N$ 
spin-orbit interaction is switched off, the $\Delta p$ splitting is almost entirely 
due to the ALS and SLS terms and is in rough agreement with the RMF values. 
By comparing the results for the ESC08c model shown in columns B and C in 
Table 2  we conclude that the magnitude of the ALS term which acts in an 
opposite way to the SLS term is about 2/3 of the SLS magnitude.

Our results demonstrate that it is highly desirable to explore further the
energy
splitting of the
$\Lambda$ $0p_{3/2}$ and $0p_{1/2}$ levels in $p$- and $sd$-shell
hypernuclei, both
experimentally and theoretically, in order to extract important
information about the spin-dependence of the $\Lambda N$ interaction, as
well as the
inner structure of the hypernuclei under study.

There are several issues left for further improvements of the present calculations. 
First, we intend to develop the code which will allow to perform calculations within an axially symmetric single particle
basis and allow for configuration mixing in the nuclear core wave function. In this case we would be able to calculate open-shell hypernuclei (such as $^{21}_{\Lambda}$Ne or 
$^{29}_{\Lambda}$Si) more precisely.
Second, we intend to implement directly the 3-body $NNN$ forces instead of
the 2-body density dependent term in our Hamiltonian. 
Third, it is desirable to incorporate the $\Lambda N$ tensor terms and explore their  
contribution to the energy splitting. 

Another extension is to include the core polarization effects. We intend to develop a scheme which couples $\Lambda$ single-particle states with one-phonon or possibly multi-phonon excitations of the core nucleus within an Equation of Motion Phonon Model (EMPM) \cite{EMPM} treating nuclear excitations in multiphonon basis. In this method the Tamm-Dancoff phonon operators $Q^{\dagger}_{\nu}$ are used to build Hilbert space spanned by one-, two- and three-phonon configurations. In this way we will get rather complex description of hypernuclei which includes not only core polarization effects but also beyond mean field correlations.

\section{Acknowledgments}

This work was partly supported by the GACR grant P203/15/04301S. Highly appreciated was the access to
computing and storage facilities provided by the MetaCentrum under the program LM2010005 and the CERIT-SC under the program Center CERIT Scientific Cloud, part of the Operational Program Research and Development for Innovations, Reg. no. CZ.1.05/3.2.00/08.0144. 
P.~V. thanks RIKEN for the kind hospitality during his stay. This work was partly supported by
RIKEN iTHES Project.

\end{document}